\documentclass[iop]{emulateapj}
\usepackage{epsfig}
\usepackage{apjfonts}
\usepackage{aas_macros}
\usepackage{amsmath}
\usepackage{lineno}

\begin{document}
\begin{linenomath*}
\title{The allowed parameter space of a long-lived neutron star as the merger remnant of GW170817}
\author{Shunke Ai$^{1}$, He Gao$^{1}$, Zi-Gao Dai$^{2,3}$, Xue-Feng Wu$^{4}$, Ang Li$^{5}$, and Bing Zhang$^{6,7,8}$}
\affiliation{
$^1$Department of Astronomy, Beijing Normal University, Beijing 100875, China; gaohe@bnu.edu.cn\\
$^2$School of Astronomy and Space Science, Nanjing University, Nanjing 210093, China; dzg@nju.edu.cn\\
$^3$Key Laboratory of Modern Astronomy and Astrophysics (Nanjing University), Ministry of Education, China\\
$^4$Purple Mountain Observatory, Chinese Academy of Sciences, Nanjing 210008, China; xfwu@pmo.ac.cn\\
$^5$Department of Astronomy, Xiamen University, Xiamen, Fujian 361005, China; liang@xmu.edu.cn\\
$^6$Department of Physics and Astronomy, University of Nevada Las Vegas, Las Vegas, NV 89154, USA; zhang@physics.unlv.edu\\
$^7$Department of Astronomy, School of Physics, Peking University, Beijing 100871, China\\
$^8$Kavli Institute of Astronomy and Astrophysics, Peking University, Beijing 100871, China
}

\begin{abstract}

Limited by the sensitivities of the current gravitational wave (GW) detectors, the central remnant of the binary neutron star (NS) merger associated with GW170817 remains an open question. Considering the relatively large total mass, it is generally proposed that the merger of GW170817 would lead to a shortly lived hypermassive NS or directly produce a black hole (BH). There is no clear evidence to support or rule out a long-lived NS as the merger remnant. Here we utilize the GW and electromagnetic (EM) signals to comprehensively investigate the parameter space that allows a long-lived NS to survive as the merger remnant of GW170817. We find that for some stiff equations of state, the merger of GW170817 could, in principle, lead to a massive NS, which has a millisecond spin period. The post-merger GW signal could hardly constrain the ellipticity of the NS. If the ellipticity reaches $10^{-3}$, in order to be compatible with the multi-band EM observations, the dipole magnetic field of the NS ($B_p$) is constrained to the magnetar level of $\sim10^{14}$ G. If the ellipticity is smaller than $10^{-4}$, $B_p$ is constrained to the level of $\sim10^{10}-10^{12}\,$G. These conclusions weakly depend on the adoption of equations of state.

\end{abstract}


\section {Introduction}
On August 17, 2017, the LIGO-Virgo scientific collaboration, for the first time, detected a gravitational wave (GW) signal from a binary neutron star (NS) merger event \citep[i.e., GW170817;][]{abbott17a}. Multi-wavelength electromagnetic (EM) counterparts to GW170817 have also been detected \cite[][for a summary]{abbott17b}.

Comprehensive analyses of the multi-messenger information have provided some important physical properties of the binary system and the merger process for GW170817. For instance, the host galaxy of the system was identified as NGC 4993 \citep{coulter17}, an early-type S0 galaxy with redshift $z=0.0097$ \citep{devaucouleurs91}. The chirp mass of the binary system is determined to be $1.188_{-0.002}^{+0.004}~{M_{\odot}}$, and the mass ratio of the two NSs was constrained to be in the range of \footnote{With the information of the optical/IR counterpart, \cite{gao17a} placed a more stringent constraint on the mass ratio of GW170817 system to the range of $0.46-0.67$.} $0.4-1.0$, so that the total mass of the system would be $2.74~M_{\odot}$ and the component mass of the binary system would be between $0.86~{M_{\odot}}$ and $2.26~{M_{\odot}}$ \citep{abbott17a}.

During the merger, a small fraction of baryonic matter is ejected, including a lanthanide-free disk wind ejecta with mass $M_{\rm ej,blue}\approx0.01-0.04~{M_{\odot}}$ and initial speed $\beta_{\rm i, blue}\approx0.2-0.3$ \citep{kasen17,cowperthwaite17,chornock17,kilpatrick17,villar17,tanvir17,gao17a}, and a lanthanide-rich dynamical ejecta\footnote{The opposite view of interpreting the blue component as due to the dynamical ejecta also exists in the literature \citep[e.g.][]{kasen17,nicholl17}.} (tidally ripped and dynamically launched matter) with mass $M_{\rm ej,red}\approx0.03-0.05~{M_{\odot}}$ and initial speed $\beta_{\rm i, red}\approx0.1-0.2$ \citep{evans17,drout17,nicholl17,smartt17,arcavi17,kasen17,cowperthwaite17,kilpatrick17,villar17,gao17a}. Such ejected matter powered an ultraviolet/optical/nearly infrared counterpart following GW170817, named AT2017gfo \cite[][for a review]{metzger17}.

The remaining matter would settle to form a new central compact object fed by an accretion disk, so that a relativistic jet was launched. When propagating through the surrounding ejecta, the jet could be structured \citep{kasliwal17,piro17,gottlieb17,lazzati17,mooley17,xiao17,zhang17,lyman18}. Internal and external dissipation of the structured jet gives rise to multi-band EM emissions, including a short duration gamma-ray burst detected by the Fermi Gamma-ray Burst Monitor (GRB 170817) \citep{goldstein17}, and late time brightening afterglow signals in the X-ray, optical and radio bands \citep{troja17,lazzati17,margutti18,troja18,meng18}. Considering that the peak isotropic luminosity of GRB 170817A ($\sim 1.7\times 10^{47}~{\rm erg\,s^{-1}}$) is abnormally low compared with other short GRBs  \citep{goldstein17,zhang17}, the late afterglow signals are relatively weak, and the rising slope of the afterglow signals is relatively small. A large binary inclination angle ($\sim20^{\circ}$) relative to our line of sight is inferred \citep{mooley17,lazzati17,lyman18}, which is well consistent with the results from the GW signal analyses \citep{abbott17b}.

What is the central remnant for GW170817 remains an open question. Considering that the total mass of
the GW170817 binary system is relatively large ($2.74~M_{\odot}$), it is generally proposed that the merger of GW170817 would lead to a temporal hypermassive NS (supported by differential rotation) which survived $10-100$ ms before collapsing into a black hole (BH) or even a BH directly \citep{margalit17,bauswein17,perego17,rezzolla18,ruiz17,ma17,metzger18}. However, our poor knowledge about the NS equation of state (EoS) makes the discussion more complex. For instance, as long as the NS EoS is stiff enough, the merger remnant of GW170817 could be a long-lived massive NS, as argued early by \cite{Dai1998a} and \cite{Dai2006}, and within such a scenario, the multi-band data of AT2017gfo could also reproduced \citep{yu17}.

In principle, post-merger GW signals could be used to probe the property of the remnant. But the search for post-merger GWs of GW170817 only provides an upper limit of the characteristic amplitude, mainly limited by the current sensitivities of the LIGO/Virgo detectors \citep{abbott17c}. In this case, we can only rely on the information of the EM signals to make constraints. It has long been proposed that when the merger remnant is a long-lived massive NS, more abundant EM signatures are expected \citep{Dai1998a,Dai1998b,Dai2004,Dai2006,zhang13,gao13,yu13,fan13,wu14,metzgerpiro14,siegel16a,siegel16b,gao15,gao17b,sun17}. For instance, after the relativistic jet propagates through the surrounding ejecta, a Poynting-flux outflow from the NS could leak out to power an extended emission through its dissipation at a large radius \citep{Dai2004,zhang13,rowlinson14,lv15,sun17}. Due to the dynamical motion of the ejecta, the ejecta materials tend to quench the outflow by closing the gap, so that the Poynting-flux outflow would be trapped inside. The outflow could then inject extra energy into the ejecta to increase its internal energy and kinetic energy, either via direct energy injection by a Poynting flux \citep{bucciantini12}, or due to heating from the bottom by the photons generated in a dissipating magnetar wind via magnetic reconnection or self-dissipation \citep{zhang13}. The heated ejecta material would power a bright thermal emission component \citep{yu13,metzgerpiro14}, normally brighter than the radioactively-driven kilo-nova \citep{li98,metzger10}. Nevertheless, the accelerated ejecta materials might also give rise to strong afterglow emission by driving a strong forward shock into an ambient medium \citep{gao13}. When the ejecta becomes optically thin, if the massive NS still exists, the dissipated photons from its Poynting-flux outflow would eventually diffuse out, resulting in a late-time re-brightening X-ray signal \citep{metzgerpiro14,gao15,gao17b,sun17}.

In this paper, we utilize the GW and electromagnetic (EM) signals to comprehensively investigate the possibility of a long-lived massive NS as the merger remnant of GW170817, and give constraints on the physical properties of the NS, if it exists.


\section{NS/QS equation of state}

For GW170817, the total gravitational mass of the binary system is estimated as $2.74M_{\odot}$ \citep{abbott17a}. Numerical simulations show that after the merger process and differential rotation phase, the mass of the uniformly rotating remnant could be estimated as \citep{hanauske17,rezzolla18}
\begin{eqnarray}
M_{\rm ur}=\delta M_g-\lambda^{-1}M_{\rm ej},
\end{eqnarray}
where $M_g$ is the initial gravitational mass of the merger remnant, $M_{\rm ej}$ is the amount of ejected baryon mass during the merger, $\delta=0.95$ is the mass fraction of the core after dynamical mass ejection, and $\lambda=1.17$ is the numerical ratio of the baryonic mass and the gravitational mass \citep{hanauske17,rezzolla18}. For GW170817, we have $M_{\rm ur}\simeq2.57M_{\odot}$, where $M_{\rm ej}\simeq0.04M_{\odot}$ is adopted.

Before the merger, the two NSs are in the Keplerian orbits, so the post-merger central remnant must be rapidly spinning. The rapid rotation could enhance the maximum gravitational mass ($M_{\rm max}$) allowed for NS survival, where $M_{\rm max}$ can be parameterized as \citep{lasky14}
\begin{eqnarray}
M_{\rm max} = M_{\rm TOV}(1+\alpha P^{\beta}),
\label{Mt1}
\end{eqnarray}
and $M_{\rm TOV}$ is the maximum NS mass for a non-rotating NS, $P$ is the spin period of the NS in units of second, and $\alpha$ and $\beta$ are functions of $M_{\rm TOV}$, NS radius ($R$), and moment of inertia ($I$).

For a given EoS, if its $M_{\rm TOV}$ is only slightly smaller than $M_{\rm ur}$, it is possible that $M_{\rm max}>M_{\rm ur}$. In this case, the uniformly rotating remnant would be a supra-massive NS. With the NS spinning down, the supra-massive NS would collapse to a BH when $M_{\rm max}$ becomes smaller than $M_{\rm ur}$. For an extremely stiff EoS, if $M_{\rm TOV}>M_{\rm ur}$, the merger remnant could even be a stable NS that never collapses.

For the purpose of this work, we adopt a series of NS EoSs with a range of maximum mass that allows $M_{\rm max}>2.57M_{\odot}$, including three new unified NS EoSs (DD2, DDME2, NL3$\omega\rho$) recently proposed \citep{fortin16}. Note that for completeness, we also consider several developed strange quark star (QS) EoSs \citep{li17}. For most of these EoSs, the numerical values for $P_k$ (Kepler period), $M_{\rm TOV}$, $R$, $I$, and the secondary parameters $\alpha$ and $\beta$ have been worked out in previous works (collected in Table 1), by using the general relativistic NS equilibrium code {\tt RNS} \citep{lasky14,li16,li17}. For the cases with $M_{\rm TOV}<2.57M_{\odot}$, we calculate their NS collapsing period ($P_{\rm col}$) by setting $M_{\rm max}=2.57M_{\odot}$. We can see that for most EoSs (except for EoS AB-L and NL3$\omega\rho$), $P_{\rm col}$ is very close to $P_k$, which are both of the order $\sim 1\,$ms.

\begin{table}
\begin{center}{\scriptsize
\caption{Basic parameters for adopted equation of states}
\begin{tabular}{cccccccc}
  \hline
 \hline
& $M_{\rm TOV}$&$R_s$&$I $ &$\alpha$ & $\beta$ &$P_K$& $P_{\rm col}$\\
&$\left(M_{\odot}\right)$&$({\rm km})$&$\left(10^{45} \rm{g~cm^2}\right)$&$\left({\rm s}^{-\beta}\right)$&~&$({\rm ms})$&$\left({\rm ms}\right)$\\
\hline
GM1 &2.37&12.05&3.33&$1.58\times 10^{-10}$&-2.84&0.72&0.85 \\
\hline
BSk21 &2.28&11.08&4.37&$2.81\times 10^{-10}$&-2.75&0.60&0.71 \\
\hline
DD2 &2.42&11.89&5.43&$1.370\times 10^{-10}$&-2.88&0.65 &0.99 \\
\hline
DDME2 &2.48&12.09&5.85&$1.966\times 10^{-10}$&-2.84&0.66 &1.20 \\
\hline
NL3$\omega\rho$ &2.75&12.99&7.89&$1.706\times 10^{-10}$&-2.88&0.69 &--\\
\hline
AB-L &2.71&13.7&4.7&$2.92\times 10^{-10}$&-2.82&0.76&--\\
\hline
CIDDM&2.09&12.43&8.645&$2.58\times 10^{-16}$&-4.93&0.83&0.93\\
\hline
CDDM1&2.21&13.99&11.67&$3.93\times 10^{-16}$&-5.00&1.00&1.20\\
\hline
CDDM2&2.45&15.76&16.37&$2.22\times 10^{-16}$&-5.18&1.12&1.70\\
\hline
MIT2 &2.08&11.48&7.881&$1.67\times 10^{-15}$&-4.58&0.71&0.81\\
\hline
MIT3 &2.48&13.71&13.43&$3.35\times 10^{-15}$&-4.60&0.85&1.50\\
\hline
\hline
 \end{tabular}
 }
\end{center}
\end{table}


\section{Constraints from the post-merger GW signal}

During the uniform rotation stage, the newly-formed NS could lose its rotation energy through both magnetic dipole radiation and GW emission \citep{shapiro83,zhang01},
\begin{eqnarray}
\dot{E}=I\Omega\dot{\Omega}=-L_{\rm sd,GW}-L_{\rm sd,EM},
\label{eq:dotE}
\label{dotE}
\end{eqnarray}
where
\begin{eqnarray}
L_{\rm sd,EM}=\frac{B_p^2 R^6 \Omega^4}{6c^3}
\end{eqnarray}
is the magnetic dipole spin-down power, and
\begin{eqnarray}
L_{\rm sd,GW}={32GI^2\epsilon^2\Omega^6\over 5c^5}
\end{eqnarray}
is the GW radiation spin-down power. $\Omega=2\pi /P$ is the angular frequency and $\dot{\Omega}$ is its time derivative, $\epsilon$ is the ellipticity of the NS, and $B_p$ is the dipolar field strength at the magnetic poles on the NS surface.

The characteristic amplitude of GWs from a rotating NS can be estimated as \citep{corsi09}
\begin{eqnarray}
h_c=fh(t)\sqrt{dt\over df},
\end{eqnarray}
where
\begin{eqnarray}
h(t)={4G\Omega^2\over c^4d}I\epsilon,
\end{eqnarray}
with $f=\Omega/ \pi$ representing the frequency of GW signals.

For a millisecond rotation NS, the spin down process could be dominated by the GW radiation, as long as $\epsilon$ is large enough. In this case, we have $\dot{E}=-{32GI^2\epsilon^2\Omega^6\over 5c^5}$. Thus, we obtain
\begin{eqnarray}
h_c=\sqrt{5IG\over Pc^3 d^2}.
\end{eqnarray}

The observation of GWs from the post-merger remnant by the LIGO-Virgo collaboration has given an upper limit strain as $h^{50\%}_{\rm rss}=5.9\times 10^{-22}{\rm Hz}^{-1/2}$ for a bar-mode model. With the definition of $h_{\rm rss}$ as
\begin{eqnarray}
h_{\rm rss}=\left[2\int_{f_{\rm min}}^{f_{\rm max}}\left(|\widetilde h_{+}(f)|^2+|\widetilde h_{\times}(f)|^2\right)df\right]^{1/2},
\end{eqnarray}
the relation between $h_c$ and $h_{\rm rss}$ can be roughly derived by\footnote{Note that in real GW data analyses, the estimation of $h_{\rm rss}$ should be much more complected. The analytical derivation here is only valid in order of magnitude, but is good enough for the purpose of this work.}
\begin{eqnarray}
h_c(\bar f)=h_{\rm rss}{\bar f \over \sqrt{2(f_{\rm max}-f_{\rm min})}},
\label{eq:hcbarf}
\end{eqnarray}
where $\bar f={(f_{\rm max}+f_{\rm min})/ 2}$ is the average value of the GW frequency and $f_{\rm max}$ and $f_{\rm min}$ are corresponding to the Kepler period $P_k$ and $P_{\rm col}$, respectively.

With equations \ref{eq:dotE} to \ref{eq:hcbarf}, we derive the upper limits of $h_c(\bar f)$ from the GW observation and the theoretical values of $h_c({\bar f})$ for the adopted EoSs, assuming that the NS spin down is dominated by the GW radiation. The inferred results are shown in Table 2. We find that for all the adopted EoSs, the theoretical value of $h_c({\bar f})$ is about one order of magnitude smaller than the observational upper limit, which indicates that even if the merger remnant of GW170817 is a millisecond massive NS, and the rotation energy of the NS is taken away by the GW radiation, the post merger GW signal is undetectable. GW observations cannot help to differentiate which power dominates the NS spin down process, nor make any constraints on the ellipticity ($\epsilon$) of the nascent NS. In the following, we separately discuss different situations with $\epsilon$ ranging from $10^{-7}$ to $10^{-3}$.

\begin{table*}
\begin{center}{\scriptsize
\caption{The characteristic amplitude of GW radiation $h_c$ and its upper limit}
\begin{tabular}{cccccccccccc}
\hline
\hline
&GM1&BSk21& DD2 &DDME2 & NL3$\omega\rho$&AB-L&CIDDM&CDDM1&CDDM2&MIT2&MIT3\\
\hline
$h_c(\bar f)(10^{-21})$&1.861&2.329&2.366&2.347&2.15&1.578&2.824&2.945&3.130&2.900&3.176\\
\hline
$h_{\rm c,upper}(\bar f)(10^{-20})$&7.535&7.968&4.672&3.695&1.599&1.521&8.489&6.076&3.559&8.257&3.509\\
\hline
\hline
 \end{tabular}
 }
\end{center}
\end{table*}

\section{Constraints from EM observations}
\subsection{Constraints from UV/optical/NIR observations}

If the merger remnant of GW170817 is a massive NS, the merger ejecta would be heated and accelerated by two different energy sources: r-process related radioactivity and dipole radiation from the NS.
Due to energy conservation, we have
\begin{eqnarray}
{dE\over dt}=\xi L_{\rm sd,EM}+{\cal D}^2L'_{\rm ra}-{\cal D}^2L'_e,
\label{eq:dEdt}
\end{eqnarray}
where $E$ is the total energy of the ejecta, $\xi$ represents the fraction of dipole radiation power injected into the ejecta, $L'_{\rm ra}$ is the comoving radioactive power, $L'_e$ represents the comoving emitted bolometric luminosity, and ${\cal D}=1/[\Gamma(1-\beta)]$ is the Doppler factor, where $\beta$ is the ejecta velocity in the lab frame and $\Gamma$ is the corresponding bulk Lorentz factor. Here we adopt the empirical expression for $L'_{\rm ra}$ proposed by \cite{korobkin12}
\begin{eqnarray}
L'_{\rm ra}=4\times10^{49}M_{\rm ej,-2}\left[{1\over2}-{1\over\pi}\arctan \left({t'-t'_0\over
t'_\sigma}\right)\right]^{1.3}~\rm erg~s^{-1},
\label{eq:Lrap}
\end{eqnarray}
where $t'_0 \sim 1.3$ s and $t'_\sigma \sim 0.11$ s. $L'_e$ could be estimated by
\begin{eqnarray}
L'_e=\left\{
\begin{array}{l l}
  {E'_{\rm int}c\over \tau R_{\rm ej}/\Gamma}, & \tau>1, \\
  {E'_{\rm int}c\over R_{\rm ej}/\Gamma}, &\tau<1.\\ \end{array} \right.\
  \label{eq:Lep}
\end{eqnarray}
$R_{\rm ej}$ is the radius of the ejecta in the lab frame, $\tau=\kappa (M_{\rm ej}/V')(R_{\rm ej}/\Gamma)$ is the optical depth of the ejecta with $\kappa$ being the opacity \citep{kasen10,kotera13}, and $E'_{\rm int}$ is the internal energy in the comoving frame. The evolution of $E'_{\rm int}$ could be expressed as
\begin{eqnarray}
{dE'_{\rm int}\over dt'}=\xi_t {\cal D}^{-2}L_{\rm d}+ L'_{\rm ra} -L'_{\rm e},
-\mathcal P'{dV'\over dt'},
\label{eq:Ep}
\end{eqnarray}
with the radiation dominated pressure $\mathcal P'=E'_{\rm int}/3V'$ and the thermalization coefficient $\xi_t=\xi e^{-1/\tau}$. The comoving volume evolution can be fully addressed by $dV'/dt'=4\pi R_{\rm ej}^2\beta c$ and $dR_{\rm ej}/dt=\beta c/ (1-\beta)$.

The dynamic equation for the ejecta could be expressed as \citep{yu13}
\begin{eqnarray}
{d\Gamma\over dt}={{dE\over dt}-\Gamma {\cal D}\left({dE'_{\rm int}\over
dt'}\right)-(\Gamma^2-1)c^2\left({dM_{\rm sw}\over dt}\right)\over
M_{\rm ej}c^2+E'_{\rm int}+2\Gamma M_{\rm sw}c^2},
\label{eq:Gt}
\end{eqnarray}
where $M_{\rm sw}=\frac{4\pi}{3}R_{\rm ej}^3nm_p$ is the shock swept mass of a medium with density of $n$.

With equations \ref{eq:dEdt} to \ref{eq:Gt}, one can easily solve the ejecta dynamics and the bolometric luminosity evolution of the ejecta thermal emission ($L_e$). Obviously, the increase of dipole radiation power ($L_{\rm sd,EM}$) could significantly enlarge the velocity of the ejecta ($\beta$) and enhance the peak value of $L_e$. From the spectral and photometric analyses of AT2017gfo, the peak value of $L_e$ and the corresponding ejecta velocity (around $\tau=1$) have been well constrained, i.e., $L_e\lesssim10^{42}{\rm erg~s^{-1}}$ and $\beta\lesssim0.3$ \citep{kasen17}. Given the tight allowed range of the spin period of the massive NS, the observations could place tight constraints on the dipole magnetic field of the NS ($B_p$). The results are collected in Table 3. We find that in order to be compatible with the UV/optical/NIR observations, if the  merger remnant of GW170817 is a massive NS, the dipole magnetic field of the NS should be less than $\sim10^{11}-10^{12}$ G (see Figure 1).

We next test two different cases with different fractions of the dipole radiation power injected into the ejecta ($\xi=0.1$ and $\xi=1$). We find that increasing $\xi$ by one order of magnitude could tighten the constraint on $B_p$ by a factor of 3. We also test different situations with $\epsilon$ ranging from $10^{-7}$ to $10^{-3}$. We find that as long as $\epsilon$ is smaller than $10^{-4}$, different $\epsilon$ no longer affects the constraints on $B_p$. However, when $\epsilon$ is of the order $\sim 10^{-3}$, the constraint on $B_p$ would become looser by one order of magnitude, mainly because the high GW emission power could rapidly slow down the NS and drive its collapse into a BH.

\begin{table*}
\begin{center}{\scriptsize
\caption{The constrained results on ${B_p} ({\rm G})$ from UV/optical/IR observation}
\begin{tabular}{cccccccccccc}
\hline
\hline
\multicolumn{12}{c}{$\xi=0.1$}\\
& GM1 & BSk21 & DD2 &DDME2 & NL3$\omega\rho$ & AB-L &CIDDM & CDDM1 & CDDM2 & MIT2 & MIT3\\
\hline
$\epsilon=10^{-3}$ & $3.39\times 10^{13}$ & $6.92\times 10^{13}$ & $2.00\times10^{13}$ & $1.12\times 10^{13}$& $2.00\times 10^{12}$& $1.35\times 10^{12} $ & $7.59\times 10^{13}$ & $2.95\times 10^{13}$ &$8.13\times 10^{12}$ & $9.33\times 10^{13}$ & $1.23\times 10^{13}$\\
$\epsilon=10^{-4}$ & $3.16\times 10^{11}$ & $6.46\times 10^{11}$ & $3.16\times10^{11}$ & $3.09\times 10^{11}$& $2.82\times 10^{11}$& $2.40\times 10^{11} $ & $6.92\times 10^{11}$ & $3.72\times 10^{11}$ &$3.24\times 10^{11}$ & $8.71\times 10^{11}$ & $3.39\times 10^{11}$\\
$\epsilon=10^{-5}$ & $2.40\times 10^{11}$ & $2.19\times 10^{11}$ & $2.04\times10^{11}$ & $2.04\times 10^{11}$& $1.78\times 10^{11}$& $1.86\times 10^{11} $ & $2.95\times 10^{11}$ & $2.95\times 10^{11}$ &$2.63\times 10^{11}$ & $2.75\times 10^{11}$ & $2.29\times 10^{11}$\\
$\epsilon=10^{-6}$ & $2.40\times 10^{11}$ & $2.14\times 10^{11}$ & $2.04\times10^{11}$ & $2.00\times 10^{11}$& $1.78\times 10^{11}$& $1.82\times 10^{11} $ &$2.88\times 10^{11}$ & $2.95\times 10^{11}$ &$2.63\times 10^{11}$ & $2.69\times 10^{11}$ & $2.29\times 10^{11}$\\
$\epsilon=10^{-7}$ & $2.40\times 10^{11}$ & $2.14\times 10^{11}$ & $2.04\times10^{11}$ & $2.00\times 10^{11}$& $1.78\times 10^{11}$& $1.82\times 10^{11} $ &$2.88\times 10^{11}$ & $2.95\times 10^{11}$ &$2.63\times 10^{11}$ & $2.69\times 10^{11}$ & $2.29\times 10^{11}$\\
  \hline
 \multicolumn{12}{c}{$\xi=1$}\\
& GM1 & BSk21 & DD2 &DDME2 & NL3$\omega\rho$ & AB-L &CIDDM & CDDM1 & CDDM2 & MIT2 & MIT3\\
  \hline
$\epsilon=10^{-3}$ & $1.07\times 10^{13}$ & $2.19\times 10^{13}$ & $6.31\times10^{12}$ & $3.55\times 10^{12}$&  $6.31\times 10^{11}$& $4.27\times 10^{11} $ & $2.40\times 10^{13}$ & $9.44\times 10^{12}$ &$2.57\times 10^{12}$ & $2.95\times 10^{13}$ & $3.89\times 10^{12}$\\
$\epsilon=10^{-4}$ & $1.00\times 10^{11}$ & $2.04\times 10^{11}$ & $1.00\times10^{11}$ & $9.77\times 10^{10}$& $8.91\times 10^{10}$ & $7.59\times 10^{10} $ & $2.19\times 10^{11}$ & $1.17\times 10^{11}$ &$1.02\times 10^{11}$ & $2.75\times 10^{11}$ & $1.07\times 10^{11}$\\
$\epsilon=10^{-5}$ & $7.59\times 10^{10}$ & $6.92\times 10^{10}$ & $6.46\times10^{10}$ & $6.46\times 10^{10}$& $5.62\times 10^{10}$& $5.89\times 10^{10} $ & $9.33\times 10^{10}$ & $9.33\times 10^{10}$ &$8.32\times 10^{10}$ & $8.71\times 10^{10}$ & $7.24\times 10^{10}$\\
$\epsilon=10^{-6}$ & $7.59\times 10^{10}$ & $6.76\times 10^{10}$ & $6.46\times10^{10}$ & $6.31\times 10^{10}$& $5.62\times 10^{10}$& $5.75\times 10^{10} $ & $9.12\times 10^{10}$ & $9.33\times 10^{10}$ &$8.32\times 10^{10}$ & $8.51\times 10^{10}$ & $7.24\times 10^{10}$\\
$\epsilon=10^{-7}$ & $7.59\times 10^{10}$ & $6.76\times 10^{10}$ & $6.46\times10^{10}$ & $6.31\times 10^{10}$& $5.62\times 10^{10}$& $5.75\times 10^{10} $ & $9.12\times 10^{10}$ & $9.33\times 10^{10}$ &$8.32\times 10^{10}$ & $8.51\times 10^{10}$ & $7.24\times 10^{10}$\\
  \hline
 \hline
 \end{tabular}
 }
\end{center}
\end{table*}

\subsection{Constraints from $\gamma$-ray and X-ray observations}

If the merger remnant of GW170817 is a massive NS, after the relativistic jet punching through the ejecta shell, the Poynting-flux outflow from the NS could leak out to power an extended emission. For GRB 170817A, \cite{zhang17} conducted a search of extended emission before and after the trigger time, which leads to a negative result. In this case, the $\gamma$-ray luminosity powered by the NS wind dissipation should not be larger than the luminosity of GRB 170817A, i.e.,  $\eta_{\gamma}L_{\rm sd,EM}\lesssim 1.7\times 10^{47} {\rm erg~s^{-1}}$.

On the other hand, when the ejecta becomes transparent, if the central NS has not collapsed, the dissipated photons from the NS wind would eventually diffuse out. Late time X-ray observations could serve as the upper limit of the X-ray luminosity powered by the NS wind dissipation, i.e.,  $\eta_x L_{\rm sd,EM}(t) e^{-\tau}\lesssim L_{X}(t)$. Here we take the X-ray data (including upper limit) from \cite{troja17} to make constraints on the dipole magnetic field of the central NS. The constrained results for different EoSs have been collected in Table 4. We find that in order to be compatible with the $\gamma$-ray and X-ray observations, if the merger remnant of GW170817 is a massive NS, the dipole magnetic field of the NS should be less than $\sim10^{11}-10^{12}$ G (see Figure 1), similar to the constraints from UV/optical/NIR data.

Here we adopt a relatively small efficiency factor ($\eta_X=10^{-4}$) to convert the spin-down luminosity to the observed X-ray luminosity \citep{xiao2017a}, which is much smaller than the inferred value from previous investigations for a sample of short GRB X-ray plateau data \citep{zhang13,rowlinson14,lv15,gao16}. Adopting a larger value of $\eta_X$ would lead to much tighter constraints on $B_p$, which may fall below $10^{10}$ G. Note that since the observed $\gamma$-ray luminosity is much larger than the X-ray data, even assuming $\eta_X=10^{-4}$ and $\eta_{\gamma}=1$, the constraints on $B_p$ are mainly from X-ray observation instead of $\gamma$-ray.

Here we only consider the $\xi=0.1$ case\footnote{In the $\xi \sim 1$ case, no constraint on the spin down luminosity could be made from X-ray and $\gamma$-ray observations, since almost all the spin-down power has been injected into the ejecta.}.  We test different situations with $\epsilon$ ranging from $10^{-7}$ to $10^{-3}$. Similar to the constrained results from UV/optical/NIR data, we find that different $\epsilon$ values do not affect the constrained results significantly, unless the GW emission power is large enough and rapidly drive the NS to collapse into a BH (when $\epsilon$ is around $10^{-4}$ or $10^{-3}$). In these cases, the constraint on $B_p$ becomes much looser since it is solely based on the $\gamma$-ray data.

\begin{table*}
\begin{center}{\scriptsize
\caption{The constrained results on ${B_p}$ from X-ray/$\gamma$-ray observations}
\begin{tabular}{cccccccccccc}
\hline
\hline
\multicolumn{12}{c}{$\xi=0.1$}\\
& GM1 & BSk21 & DD2 &DDME2 & NL3$\omega\rho$ &AB-L& CIDDM & CDDM1 & CDDM2 & MIT2 & MIT3 \\
\hline
$\epsilon=10^{-3}$ & $4.15\times 10^{13}$ & $3.71\times 10^{13}$ & $3.52\times10^{13}$ & $3.45\times 10^{13}$&  $6.76\times 10^{12}$& $4.47\times 10^{12}$ & $5.02\times 10^{13}$ & $5.12\times 10^{13}$ &$4.49\times 10^{13}$ & $4.67\times 10^{13}$ & $3.93\times 10^{13}$\\
$\epsilon=10^{-4}$ & $2.82\times 10^{12}$ & $3.71\times 10^{13}$ & $2.63\times10^{12}$ & $6.17\times 10^{11}$& $7.24\times 10^{11}$& $5.13\times 10^{11}$ & $5.02\times 10^{13}$ & $3.47\times 10^{12}$ &$5.37\times 10^{11}$ & $4.67\times 10^{13}$ & $6.61\times 10^{11}$\\
$\epsilon=10^{-5}$ & $2.88\times 10^{11}$ & $2.63\times 10^{11}$ & $2.52\times10^{11}$ & $2.45\times 10^{11}$& $2.57\times 10^{11}$& $2.51\times 10^{11}$ & $3.55\times 10^{11}$ & $3.55\times 10^{11}$ &$3.09\times 10^{11}$ & $3.31\times 10^{11}$ & $2.75\times 10^{11}$\\
$\epsilon=10^{-6}$ & $2.88\times 10^{11}$ & $2.57\times 10^{11}$ & $2.45\times10^{11}$ & $2.40\times 10^{11}$& $2.29\times 10^{11}$& $2.40\times 10^{11}$ & $3.47\times 10^{11}$ & $3.55\times 10^{11}$ &$3.09\times 10^{11}$ & $3.23\times 10^{11}$ & $2.69\times 10^{11}$\\
$\epsilon=10^{-7}$ & $2.88\times 10^{11}$ & $2.57\times 10^{11}$ & $2.45\times10^{11}$ & $2.40\times 10^{11}$& $2.29\times 10^{11}$& $2.34\times 10^{11}$ &$3.47\times 10^{11}$ & $3.55\times 10^{11}$ &$3.09\times 10^{11}$ & $3.23\times 10^{11}$ & $2.69\times 10^{11}$\\
 \hline
 \hline
 \end{tabular}
 }
\end{center}
\end{table*}


\subsection{Constraints from radio observations}

Energy injection from the central NS, if exists, could significantly accelerate the ejecta. As long as the kinetic energy of the ejecta is large enough, the forward shock into the ambient medium could give rise to strong afterglow emission, at least in the radio band. The afterglow emission should not outshine the late radio observations, which could make further constraints on the dipole magnetic field of the central NS.

Recently, \cite{mooley17} has applied the ejecta-medium forward shock model to interpret the data of radio counterpart of GW170817, and they find that the radio light curve could be well fitted within such a model, as long as the kinetic energy of ejecta could reach $5\times 10^{50}{\rm ergs}$ and the density of interstellar medium is $n=0.03 {\rm cm^{-3}}$.

Solving the dynamical evolution of the ejecta with medium density as $n=0.03 {\rm cm^{-3}}$, we can make constraints on $B_p$ by setting $5\times 10^{50}{\rm ergs}$ as the upper limit of the ultimate ejecta kinetic energy. The constrained results are shown in Table 5. We find that in order to be compatible with the radio observations, if the merger remnant of GW170817 is a massive NS, the dipole magnetic field of the NS should be less than $\sim10^{12}-10^{14}$\,G (see Figure 1).

Here we consider two cases with $\xi=0.1$ and $\xi=1$. Again, we find that increasing $\xi$ by one order of magnitude could tighten the constraint on $B_p$ by a factor of 3. We also test different situations with $\epsilon$ values. We find that when $\epsilon$ is in order of $10^{-3}$, the $B_p$ upper limit is of the order $\sim10^{14}$ G. When $\epsilon=10^{-4}$, the $B_p$ upper limit is of the order $\sim 10^{13}$ G. As long as $\epsilon$ is equaling to or less than $10^{-5}$, different $\epsilon$ no longer affects the constrained results on $B_p$, and the $B_p$ upper limit is of the order $\sim 10^{12}$ G.

\begin{table*}
\begin{center}{\scriptsize
\caption{The constrained results on ${B_p}$ from radio observation}
\begin{tabular}{cccccccccccc}
\hline
\hline
 \multicolumn{12}{c}{$\xi=0.1$}\\
& GM1 & BSk21 & DD2 & DDME2 & NL3$\omega\rho$ & AB-L & CIDDM & CDDM1 & CDDM2 & MIT2 & MIT3  \\
\hline
$\epsilon=10^{-3}$ & $2.45\times 10^{14}$ & $3.24\times 10^{14}$ & $1.82\times10^{14}$ & $1.51\times 10^{14}$ & $6.92\times 10^{13}$& $4.90\times 10^{13}$ & $3.89\times 10^{14}$ & $2.51\times 10^{14}$ & $1.35\times 10^{14}$ &$4.27\times10^{15}$ & $1.62\times 10^{14}$\\
$\epsilon=10^{-4}$ & $3.02\times 10^{13}$ & $3.63\times 10^{13}$  & $2.29\times10^{13}$ & $1.74\times 10^{13}$& $9.33\times 10^{12}$& $6.61\times 10^{12}$ & $4.47\times10^{13}$ & $3.24\times 10^{13}$ & $1.51\times 10^{13}$ &$4.68\times10^{13}$ & $1.91\times 10^{13}$\\
$\epsilon=10^{-5}$ & $2.45\times 10^{12}$ & $3.31\times 10^{12}$ & $2.00\times10^{12}$ & $1.95\times 10^{12}$& $1.78\times 10^{12}$& $1.55\times 10^{12}$ & $3.98\times 10^{12}$ & $2.51\times 10^{12}$ & $2.09\times 10^{12}$ & $4.37\times 10^{12}$ & $2.14\times 10^{12}$\\
$\epsilon=10^{-6}$ & $1.62\times 10^{12}$ & $1.48\times 10^{12}$ & $1.38\times10^{12}$ & $1.35\times 10^{12}$& $1.20\times 10^{12}$ & $1.26\times 10^{12}$ & $2.00\times 10^{12}$ & $2.00\times 10^{12}$ &$1.78\times 10^{12}$ & $1.82\times 10^{12}$ & $1.55\times 10^{12}$\\
$\epsilon=10^{-7}$ & $1.62\times 10^{12}$ & $1.45\times 10^{12}$ & $1.38\times10^{12}$ & $1.35\times 10^{12}$& $1.20\times 10^{12}$& $1.26\times 10^{12}$ & $1.95\times 10^{12}$ & $2.00\times 10^{12}$ &$1.78\times 10^{12}$ & $1.82\times 10^{12}$ & $1.55\times 10^{12}$\\
\hline
\multicolumn{12}{c}{$\xi=1$}\\
& GM1 & BSk21 & DD2 &DDME2 & NL3$\omega\rho$ & AB-L & CIDDM & CDDM1 & CDDM2 & MIT2 & MIT3 \\
\hline
$\epsilon=10^{-3}$ & $7.41\times 10^{13}$ & $1.02\times 10^{14}$  & $5.75\times10^{13}$ & $4.68\times 10^{13}$&  $2.14\times 10^{13}$ & $1.51\times 10^{13}$ & $1.20\times10^{14}$ & $7.76\times 10^{13}$ & $4.27\times 10^{13}$ &$1.32\times10^{14}$ & $5.13\times 10^{13}$\\
$\epsilon=10^{-4}$ & $9.12\times 10^{13}$ & $1.12\times 10^{13}$  & $7.08\times10^{12}$ & $5.50\times 10^{12}$&  $2.88\times 10^{12}$& $2.04\times 10^{12}$ & $1.38\times10^{13}$ & $9.77\times 10^{12}$ & $4.79\times 10^{12}$ &$1.44\times10^{13}$ & $6.03\times 10^{12}$\\
$\epsilon=10^{-5}$ & $7.41\times 10^{11}$ & $1.02\times 10^{12}$ & $6.17\times10^{11}$ & $6.17\times 10^{11}$&  $5.50\times 10^{11}$& $4.78\times 10^{11}$ & $1.23\times 10^{12}$ & $7.76\times 10^{11}$ & $6.61\times 10^{11}$ & $1.35\times 10^{12}$ & $6.76\times 10^{11}$\\
$\epsilon=10^{-6}$ & $5.13\times 10^{11}$ & $4.57\times 10^{11}$ & $4.37\times10^{11}$ & $4.27\times 10^{11}$&  $3.80\times 10^{11}$& $3.89\times 10^{11}$ & $6.17\times 10^{11}$ & $6.31\times 10^{11}$ &$5.50\times 10^{11}$ & $5.75\times 10^{11}$ & $4.90\times 10^{11}$\\
$\epsilon=10^{-7}$ & $5.13\times 10^{11}$ & $4.57\times 10^{11}$ & $4.37\times10^{11}$ & $4.27\times 10^{11}$&  $3.72\times 10^{11}$& $3.89\times 10^{11}$ & $6.17\times 10^{11}$ & $6.31\times 10^{11}$ &$5.50\times 10^{11}$ & $5.75\times 10^{11}$ & $4.90\times 10^{11}$\\
 \hline
 \hline
 \end{tabular}
 }
\end{center}
\end{table*}

\begin{figure*}[ht!]
\begin{center}
\begin{tabular}{ll}
\resizebox{90mm}{!}{\includegraphics[]{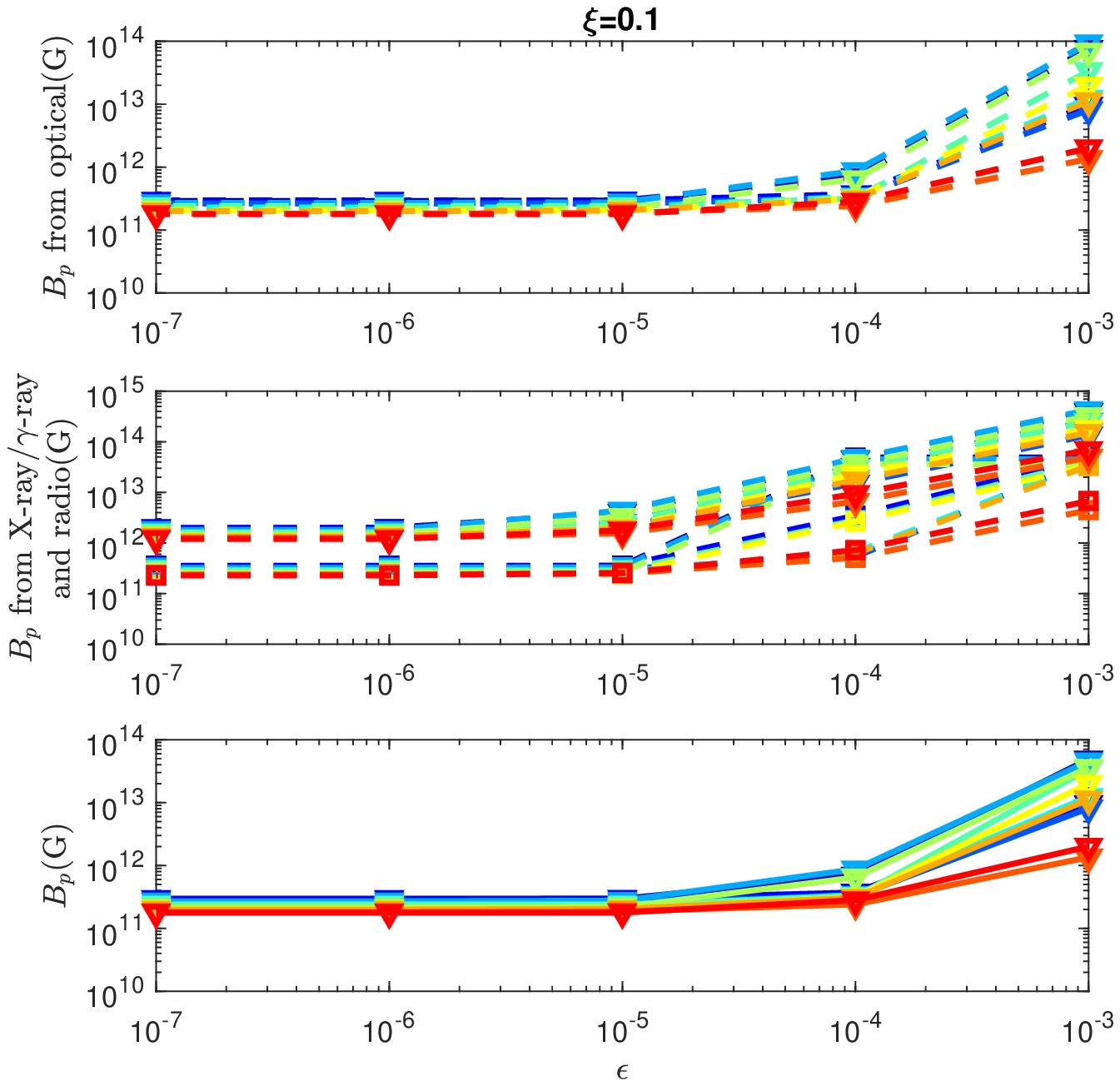}} &
\resizebox{90mm}{!}{\includegraphics[]{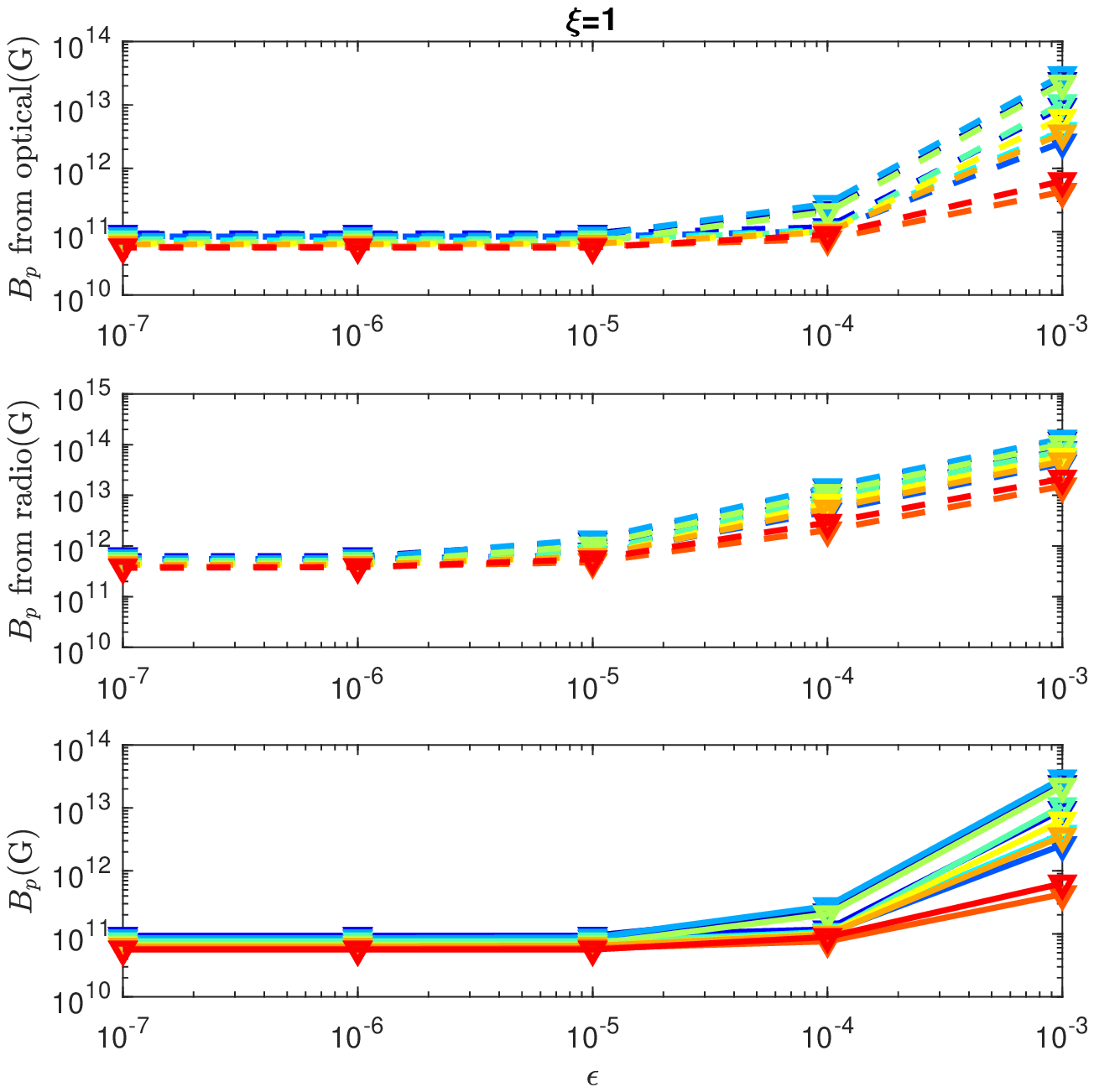}}
\end{tabular}
\caption{Constraints on the dipole magnetic field of the central NS from multi-band observations. The left panels are for $\xi=0.1$ situation and the right panels are for $\xi=1$ situation. The top panels show constrained results from UV/optical/NIR data. Constrained results from $\gamma$-ray/X-ray (marked with squares) and radio (marked with triangles) data are shown in the middle panels. The bottom panels show the final results from multi-band data. Different colors correspond to different EoSs.}
\label{fig:fit}
\end{center}
\end{figure*}

\subsection{Summary of quantitative constraints}

Combining all the constraints from GW and multi-band EM observations, we find that if the merger remnant of GW170817 is a massive NS, the NS should have millisecond spin period, but relatively low dipole magnetic field ($B_p$ as low as $\sim10^{11}$ G). The ellipticity of the NS is hardly constrained. If the ellipticity could reach $10^{-3}$, $B_p$ is constrained to the level of $\sim10^{14}$ G. Otherwise, $B_p$ is limited to the level of $\sim10^{10}-10^{12}$ G. These conclusions weakly depend on the adoption of NS EoSs.  Specifically, for GM1, the upper limit of spin period is 0.85{\rm ms} and the upper limit of $B_p$ is $7.59\times 10^{10} {\rm G}$. For Bsk21, the upper limit of spin period is 0.71{\rm ms} and the upper limit of $B_p$ is $6.76\times 10^{10} {\rm G}$. For DD2, the upper limit of spin period is 0.99{\rm ms} and the upper limit of $B_p$ is $6.46\times 10^{10} {\rm G}$. For DDME2, the upper limit of spin period is 1.20{\rm ms} and the upper limit of $B_p$ is $6.31\times 10^{10} {\rm G}$. For NL3$\omega\rho$, no upper limit of spin period could be given and the upper limit of $B_p$ is $5.62\times 10^{10} {\rm G}$. For AB-L, no upper limit of spin period could be given and the upper limit of $B_p$ is $5.75\times 10^{10} {\rm G}$. For CIDDM, the upper limit of spin period is 0.93{\rm ms} and the upper limit of $B_p$ is $9.12\times 10^{10} {\rm G}$. For CDDM1, the upper limit of spin period is 1.20{\rm ms} and the upper limit of $B_p$ is $9.33\times 10^{10} {\rm G}$. For CDDM2, the upper limit of spin period is 1.70{\rm ms} and the upper limit of $B_p$ is $8.32\times 10^{10} {\rm G}$. For MIT2, the upper limit of spin period is 0.81{\rm ms} and the upper limit of $B_p$ is $8.51\times 10^{10} {\rm G}$. For MIT3, the upper limit of spin period is 1.50{\rm ms} and the upper limit of $B_p$ is $7.24\times 10^{10} {\rm G}$. Final constrained results are collected in Table 6.

\begin{table*}
\begin{center}{\scriptsize
\caption{ Final constrained results on ${B_p} ({\rm G})$ }
\begin{tabular}{cccccccccccc}
\hline
\hline
\multicolumn{12}{c}{$\xi=0.1$}\\
& GM1 & BSk21 & DD2 &DDME2 & NL3$\omega\rho$ & AB-L &CIDDM & CDDM1 & CDDM2 & MIT2 & MIT3\\
\hline
$\epsilon=10^{-3}$ & $3.39\times 10^{13}$ & $3.71\times 10^{13}$ & $2.00\times10^{13}$ & $1.12\times 10^{13}$& $2.00\times 10^{12}$& $1.35\times 10^{12} $ & $5.02\times 10^{13}$ & $2.95\times 10^{13}$ &$8.13\times 10^{12}$ & $4.67\times 10^{13}$ & $1.23\times 10^{13}$\\
$\epsilon=10^{-4}$ & $3.16\times 10^{11}$ & $6.46\times 10^{11}$ & $3.16\times10^{11}$ & $3.09\times 10^{11}$& $2.82\times 10^{11}$& $2.40\times 10^{11} $ & $6.92\times 10^{11}$ & $3.72\times 10^{11}$ &$3.24\times 10^{11}$ & $8.71\times 10^{11}$ & $3.39\times 10^{11}$\\
$\epsilon=10^{-5}$ & $2.40\times 10^{11}$ & $2.19\times 10^{11}$ & $2.04\times10^{11}$ & $2.04\times 10^{11}$& $1.78\times 10^{11}$& $1.86\times 10^{11} $ & $2.95\times 10^{11}$ & $2.95\times 10^{11}$ &$2.63\times 10^{11}$ & $2.75\times 10^{11}$ & $2.29\times 10^{11}$\\
$\epsilon=10^{-6}$ & $2.40\times 10^{11}$ & $2.14\times 10^{11}$ & $2.04\times10^{11}$ & $2.00\times 10^{11}$& $1.78\times 10^{11}$& $1.82\times 10^{11} $ & $2.88\times 10^{11}$ & $2.95\times 10^{11}$ &$2.63\times 10^{11}$ & $2.69\times 10^{11}$ & $2.29\times 10^{11}$\\
$\epsilon=10^{-7}$ & $2.40\times 10^{11}$ & $2.14\times 10^{11}$ & $2.04\times10^{11}$ & $2.00\times 10^{11}$& $1.78\times 10^{11}$& $1.82\times 10^{11} $ & $2.88\times 10^{11}$ & $2.95\times 10^{11}$ &$2.63\times 10^{11}$ & $2.69\times 10^{11}$ & $2.29\times 10^{11}$\\
  \hline
 \multicolumn{12}{c}{$\xi=1$}\\
& GM1 & BSk21 & DD2 &DDME2 & NL3$\omega\rho$ & AB-L &CIDDM & CDDM1 & CDDM2 & MIT2 & MIT3\\
  \hline
$\epsilon=10^{-3}$ & $1.07\times 10^{13}$ & $2.19\times 10^{13}$ & $6.31\times10^{12}$ & $3.55\times 10^{12}$& $6.31\times 10^{11}$& $4.27\times 10^{11} $ & $2.40\times 10^{13}$ & $9.44\times 10^{12}$ &$2.57\times 10^{12}$ & $2.95\times 10^{13}$ & $3.89\times 10^{12}$\\
$\epsilon=10^{-4}$ & $1.00\times 10^{11}$ & $2.04\times 10^{11}$ & $1.00\times10^{11}$ & $9.77\times 10^{10}$& $8.91\times 10^{10}$& $7.59\times 10^{10} $ & $2.19\times 10^{11}$ & $1.17\times 10^{11}$ &$1.02\times 10^{11}$ & $2.75\times 10^{11}$ & $1.07\times 10^{11}$\\
$\epsilon=10^{-5}$ & $7.59\times 10^{10}$ & $6.92\times 10^{10}$ & $6.46\times10^{10}$ & $6.46\times 10^{10}$& $5.62\times 10^{10}$& $5.89\times 10^{10} $ & $9.33\times 10^{10}$ & $9.33\times 10^{10}$ &$8.32\times 10^{10}$ & $8.71\times 10^{10}$ & $7.24\times 10^{10}$\\
$\epsilon=10^{-6}$ & $7.59\times 10^{10}$ & $6.76\times 10^{10}$ & $6.46\times10^{10}$ & $6.31\times 10^{10}$& $5.62\times 10^{10}$& $5.75\times 10^{10} $ & $9.12\times 10^{10}$ & $9.33\times 10^{10}$ &$8.32\times 10^{10}$ & $8.51\times 10^{10}$ & $7.24\times 10^{10}$\\
$\epsilon=10^{-7}$ & $7.59\times 10^{10}$ & $6.76\times 10^{10}$ & $6.46\times10^{10}$ & $6.31\times 10^{10}$& $5.62\times 10^{10}$& $5.75\times 10^{10} $ & $9.12\times 10^{10}$ & $9.33\times 10^{10}$ &$8.32\times 10^{10}$ & $8.51\times 10^{10}$ & $7.24\times 10^{10}$\\
  \hline
 \hline
 \end{tabular}
 }
\end{center}
\end{table*}

\subsection{Other constraints}

Some other information may also pose constraints to the merger product. However, since they depend on complicated physical factors, an quantitative constraint is not easy achieve. Nonetheless, it is worth discussing these factors qualitatively.

The first factor is the inferred mass and velocity of the ejecta. Fitting the optical/IR data of GW170817 led to an estimate of the mass, velocity and opacity of both the blue and red components \cite[][for a review]{metzger17}. However, it is unclear which component originates from the dynamical ejecta and which originates from a neutrino-driven wind \citep[e.g.][]{gao17a}. It was proposed \citep{metzger18} that a rapidly-spinning HMNS with an ordered surface magnetic field strength of $\sim10^{14}$ G and extended lifetime ($\sim0.1-1$ s) is required to simultaneously explain the velocity, total mass and electron fraction of the blue component. It is hard to evaluate the consequence of a long-lived pulsar on the ejecta parameters. For the long-lived pulsar parameters constrained from the above quantitative analysis, usually a neutron star with a lower $B_p$ is required. It may appear that the $B$ field is too low to accelerate the ejecta to the desired velocity. On the other hand, the longer life time of the pulsar would have a longer duration of energy injection into the ejecta, so that the fast velocity may be also achieved. A long-lived pulsar may be questioned since it may over-eject mass due to neutrino-driven wind from the surface of the neutron star. However, the neutrino cooling time scale of a new-born neutron star is typically much shorter than the spin down time scale of a low-field pulsar, so that neutrino-driven wind mass may not be significantly larger than the HMNS case.

Another constraint may come from the possible lanthanide abundance in the ejecta. The existence a significant amount of lanthanides, as evidenced by the distinct ``double-peaked'' spectrum of AT2017gfo, may disfavor a strong magnetar wind, since such a wind would deeply ionize the lanthanides so that the opacity would be greatly reduced. However, the above quantitative constraints favor a low-$B$ pulsar, so that the spin down luminosity of the pulsar would not be large enough to destroy lanthanides. The existence of lanthanides in the ejecta therefore may not pose great extra constraints to the pulsar parameters.

Finally, in the above analyses, we have taken $\epsilon$ and $B_p$ as independent parameters. In principle, magnetic distortion may play the dominant role in creating and maintaining the $\epsilon$ for a newborn millisecond NS. Previous analytical and numerical studies suggest that within the magnetic distortion scenario, $\epsilon \propto B_p^2$ is usually invoked \citep{bonazzola96b,haskell08}. Recently, an ellipticity of the order $\epsilon\sim0.005$ for a rapidly spinning (millisecond), strongly magnetized ($10^{15}$ G), supramassive NS has been inferred from the statistical observational properties of Swift SGRBs \citep{gao16}. With such a normalization, the relation between $\epsilon$ and $B_p$ can be then calibrated \citep{gao17c}. According to this relation, the above quantitative constraints for $\epsilon=10^{-3}$ would be no longer relevant, since $B_p$ is required to be in the level of $10^{15}$ G in order to achieve $\epsilon=10^{-3}$. This is inconsistent with previous constrained results ($B_p<10^{14}$ G). The quantitative constraints would be still valid if a different distortion mechanism other than magnetic distortion is at play. For low $\epsilon$ value cases, since different $\epsilon$ values no longer affect the constrained results on $B_p$, the quantitative constraints discussed above still stand.

\section{Conclusions and discussions}

The recent observations of GW170817 and its EM counterpart have opened a new era of GW-led multi-messenger astronomy. Comprehensive analyses of the multi-messenger information have provided some important physical properties of the compact objects and the merger process for GW170817. However, what is the central remnant for GW170817 remains unknown. In this paper, we have investigated the possibility of a long-lived massive NS as the merger remnant of GW170817, and given constraints on the physical properties of the NS, by invoking as much as available multi-messenger information.

We found that there is no clear exclusion for a massive NS as the merger remnant of GW170817, but the parameter space for the newborn NS is limited. Constraints from GW and multi-band EM observations show that if the merger remnant of GW170817 is a massive NS, the NS should have a millisecond spin period, and relatively low dipole magnetic field (as low as $\sim10^{11}$ G). The ellipticity of the NS is hardly constrained. If the ellipticity reaches $10^{-3}$, $B_p$ is constrained to the level of $\sim10^{14}$ G. Otherwise, $B_p$ is limited to the level of $\sim10^{10}-10^{12}$ G. The conclusions weakly depend on the adoption of NS or QS EoSs.

The constraints are mainly contributed by the UV/optical/NIR and X-ray observations. It seems that the constraints from the current radio data is looser than the results from other bands. This result is based on the assumption that the radio signal is mainly generated by the ejecta-medium forward shock. It is generally believed that the late time brightening signals in X-ray, optical and radio bands are all contributed by the external dissipation of a structured jet \citep{troja17,lazzati17,margutti18,troja18}. If this is true, or if the radio signal starts to decay sooner in the future (more than one hundred days after GW170817), or if \cite{mooley17} overestimated the kinetic energy of the ejecta, even more strict constraints could be placed on the NS properties.

In this paper, we considered two NS EoSs (AB-L and NL3$\omega\rho$) with $M_{\rm TOV}>2.57M_{\odot}$. If such EoSs are valid, the merger remnant of GW170817 would be a stable NS that never collapses to a black hole. For these cases, there is no restriction on the spin period of the NS. In principle, if the spin period is extremely large (of the order $\sim 100$\,ms), current EM observations would fail to give any constraints on the dipole magnetic field of the NS. However, since the newborn NS arises from NS-NS merger scenario, the initial spin period should be close to 1 ms. Even with an extremely large ellipticity as $\epsilon\geq0.001$, GW radiation cannot spin down the NS from 1 ms to 100 ms within a reasonable timescale. Even under these extreme EoSs, the newborn NS is hardly possible to have a large $B_p$, unless some other mechanisms \citep[e.g. r-mode instability,][]{Andersson1998, Lindblom1998, Dai2016} could somehow carry away its angular momentum at very early stage. For these two EoSs, we only consider dipole radiation and GW emission as the spin down mechanism and take the Kepler period as the initial spin period.

Previous analyses on the short GRB data indicate that the dipole magnetic field of the merger producing NS is typically large ($\sim10^{15}$ G; \cite{rowlinson14,lv15,gao16}). If the central remnant of GW170817 is a massive NS, it is an outlier compared with other cases. But it is interesting to note that compared with other short GRBs, GRB 170817A is also an outlier in terms of luminosity.

For millisecond NSs, the detection horizons for the third generation gravitational wave detectors, such as Einstein Telescope could reach 600 Mpc \citep{gao17c}. Future detections of the post-merger GW signals would be essential to determine NS-NS merger remnants, and further reveal the NS equation of state.

\acknowledgments
This work was supported by the National Basic Research Program (973 Program) of China (Grant No. 2014CB845800), the National Natural Science Foundation of China (under Grant No. 11722324, 11725314, 11603003, 11633001, 11690024, and 11573014), and the Strategic Priority Research Program of the Chinese Academy of Sciences (Grant No. XDB23040100).

\end{linenomath*}

\end{document}